\documentclass{emulateapj}

\begin{document}

\title{A High Contrast Imaging Survey of SIM Lite Planet Search Targets}

\author{Angelle M. Tanner\altaffilmark{1}, 
Christopher R. Gelino\altaffilmark{2}, 
Nicholas M. Law\altaffilmark{3}}

\altaffiltext{1}{Georgia State University, Department of Astronomy, One Park Place,  Atlanta, GA, angelle.tanner@gmail.com }
\altaffiltext{2}{Infrared Processing and Analysis Center, 770 S. Wilson Ave. Pasadena, CA 91125}
\altaffiltext{3}{Dunlap Institute for Astronomy and Astrophysics, University
of Toronto, 50 St. George Street, Toronto Ontario, Canada, M5S 3H4}

\begin{abstract}

With the development of extreme high contrast ground-based adaptive optics instruments and space missions
aimed at detecting and characterizing Jupiter- and terrestrial-mass planets, it is critical that each target star
be thoroughly vetted to determine whether it is a viable target given both the instrumental design and scientific goals of the program. 
With this in mind, we have conducted a high contrast imaging survey of mature AFGKM stars with the PALAO/PHARO instrument on the Palomar 200 inch telescope.  
The survey reached sensitivities sufficient to detect brown dwarf companions at separations of $>$ 50 AU.  
The results of this survey will be utilized both by future direct imaging projects such as GPI, SPHERE and P1640 and indirect detection missions such 
as SIM Lite.  Out of 84 targets, all but one have no close-in (0.45-1$\arcsec$)  companions and 64 (76\%) have no stars at all within the 25$\arcsec$ field-of-view. 
The sensitivity contrasts in the K$_s$ passband ranged from 4.5 to 10 for this set of observations. 
These stars were selected as the best nearby targets for habitable planet searches owing to their long-lived habitable zones ($>$ 1 billion years).
We report two stars, GJ 454 and GJ 1020, with previously unpublished proper motion companions. In both cases, the companions are
stellar in nature and are most likely M dwarfs based on their absolute magnitudes and colors. 
Based on our mass sensitivities and level of completeness, we can
place an upper limit of $\sim$17\% on the presence of brown dwarf companions with masses $>$40 M$_J$ at separations of $>$1 arcsecond.
We also discuss the importance of including statistics on those stars with no detected companions in their field of view for the sake
of future companion searches and an overall understanding of the population of low-mass objects
around nearby stars. 
\end{abstract}

\section{Introduction}

For the last decade, planet search efforts around nearby stars have been led by radial velocity
surveys which are sensitive to Jupiter mass planets with close-in ($<$5 AU) orbits (Butler et al. 2006). 
High contrast imaging surveys have recently culminated with the direct detection of 
Jupiter-mass planets at wider separations ($>$ 20 AU, Marois et al. 2008; Kalas et al. 2008). 
While the radial velocity surveys are approaching sensitivities capable of detecting hot super-Earth mass planets around solar-mass
stars and even habitable terrestrial-mass planets around some low-mass stars, Earth analogs remain out of reach due
to intrinsic stellar jitter. 
SIM Lite (formally called SIM PlanetQuest, Unwin et al. 2008) will be the first instrument capable of detecting Earth analogs in the
habitable zone of nearby stars.  SIM Lite is a planned space-based interferometer with two 50 cm telescopes separated by 6 meters. It is capable of
achieving a single measurement positional accuracy of 1 $\mu$as, enough to detect terrestrial planets in the habitable zone of nearby solar-type
stars. 

There are two SIM Key projects chosen to conduct surveys 
capable of detecting terrestrial-mass planets around nearby main sequence stars - 
Discovery of Planetary Systems (Geoff Marcy, PI) and 
the Extrasolar Planets Interferometric Survey (EPIcS, Mike Shao, PI). 
These projects will carry out an astrometric search for rocky planets around
$\sim$200 stars located within 30 pc of the Sun. 

There are many precursor programs currently being conducted to 
characterize and vet all potential targets for all the SIM Key projects (Unwin et al. 2008). These programs include radial velocity studies of
the reference stars utilized in the narrow-angle observations, accurate spectral type
determinations through optical spectroscopy, and photometry studies of the stars in
the young star planet search program. Some of these programs have resulted in planet detections themselves (Niedzielski et al. 2007)
or data that can be applied to upcoming planet surveys (Tanner et al. 2007). Here, we present the results of a
companion survey of stars included in both of the terrestrial planet SIM Lite surveys.
The observational goal of this companion survey is to identify bright, nearby companions
to the target stars. A companion with $\Delta$V $<$ 4 and within 1$\arcsec$ can bias the position of the photocenter
of the star, thus reducing the astrometric accuracy. While this survey was designed as a precursor program for
SIM Lite, the observations presented here also serve as reconnaissance  
data for upcoming high contrast imaging surveys such as the Gemini Planet Imager (GPI) (Macintosh et al. 2006), the VLT SPHERE 
coronagraph (Beuzit et al. 2008) and P1640, the recently commissioned
Lyot AO coronagraph on the Palomar 200 inch telescope (Hinkley et al. 2008).
As a result of the high contrast images collected for this survey, for most
of the stars in the sample,  we are sensitive to brown dwarf mass objects at separations of $>$ 50 AU. 
Because this program is sensitive to brown dwarf companions at wide
separations, the results serve as an additional probe of the brown dwarf desert.

Section 2 describes the criteria by which we created the sample for the survey, $\S$ 3 details our observations, $\S$ 4 describes the
data reduction and analysis, $\S$ 5 summarizes the results of the survey and our achieved sensitivities and in $\S$ 6 we compare our results to
previous surveys and discuss the importance of publishing non-detections. 

\section{Sample Selection} 

The two SIM Lite terrestrial planet search 
programs coordinated their target selection so as to create unique lists for each team.
Since one of the projects is primarily interested in finding terrestrial planets
in the habitable zone (``Extrasolar Planets Interferometric Survey (EPIcS)'', Shao, PI) and the other is aimed at detecting 
all planets around stars within 8 pc  (``Discovery of Planetary Systems'', Marcy, PI), it was not difficult to 
agree on the two samples.\footnote{The highest priority members of the combined target list are listed at http://www.physics.sfsu.edu/$\sim$chris/SIM/t1.html}
Target stars were selected based on spectral type, heliocentric distance, 
brightness, companion separation (if applicable), habitable zone location, and the orbital period at the habitable zone.

Since these lists were created, the EPIcS program completed a study on
how to further optimize the list. An additional sample list of 240 stars has been created 
for the purpose of performing Monte Carlo simulations to predict the potential yield of 
a SIM Lite terrestrial-planet search program (Catanzarite et al. 2006). This
SIM-optimized target list is derived from an initial list
of 2350 stars taken from the Hipparcos catalog, with distances
of less than 30 pc (Turnbull \& Tarter 2003). They excluded stars
with luminosity greater than 25 times solar, thereby eliminating
giants from our sample. To eliminate the possibility of fringe
contamination from a binary companion, we applied the following:
stars with a known companion closer than 0.4$\arcsec$ and $\Delta$V $<$ 4 were excluded.
If the target-star candidate had a wide
binary companion that was separated by more than 1.5$\arcsec$, the
companion was added to the list of target-star candidates.
Further improvements on the target ranking 
will include considerations of stellar metallicity and the number and 
orientation of the reference stars with respect to the target. In the end, 
these stars will represent the best targets for a micro-arcsecond astrometric planet search. 

After two years of observations, we have collected observations for
a subset of 84 nearby, main sequence (V) stars taken from both the original EPIcS sample and the
one created for Catanzarite et al. (2006). The properties of each of the stars in the sample are given in Table~\ref{sample}.
The stars in our sample have distances of 3.5-30 pc, spectral types of M4-A3V and V magnitudes
of 0.77-11.35.  Most if not all of the stars in the sample are mature stars with ages of 1-10 Gyr. We made sure not to 
reobserve those stars with published AO coronagraphic or Hubble Space telescope observations  (Carson et al. 2007; Lowrance et al. 2005; Metchev et al. 2005).

\section{Observations}

Observations were obtained with the Palomar Observatory Hale 5m Telescope using the PHARO near-IR camera (Hayward et al. 2001) 
behind the PALAO adaptive optics (AO) system (Troy et al. 2000).  Observing dates and sky conditions are given in Table 2.  We used a 25 
mas/pixel scale camera (25$\arcsec$ field of view) and the 0.45$\arcsec$ radius occulting spot.  Each star was observed in the K$_s$ filter 
(2.16 $\micron$) with integration times of 1--30 seconds, depending on the brightness of the star.  Stacks of 20-100
short exposure images were collected for each star resulting in effective integration times of 100-1200 seconds\footnote{Since the original SIM Lite Palomar survey started in 2004 was meant as a snap-shot program intended to look for close, stellar binaries, a subset
of the coronagraphic observations were performed with a neutral density filter which reduced the effective integration times by a factor of 100. In most cases, those stars were reobserved without the neutral density filter to bring the effective integration times up to the level of the rest of the survey targets.}. 
Stacks of sky images  were taken adjacent to each set of target images by offsetting 
30$\arcsec$ from the target in the four cardinal directions and turning off the AO system. 
For flux calibration, observations of the target stars were taken by inserting the neutral density filter,  
offsetting the star from the coronagraph and placing it in a five point dither pattern to  
allow for adequate sky subtraction\footnote{Flux calibration images were not 
collected during the June 2004 observing run.}. These offset images allowed us to determine that the FWHM of the AO corrected observations varied from 0.2 to 1.3$\arcsec$
with Strehl ratios of 0.20-0.74. The seeing quality varies considerably for the different
observing runs, from ``good'' (0.5$\arcsec$) to bad ($>$3$\arcsec$) based on the FWHM of the point spread function (PSF) with no AO correction (see Table~\ref{obs}). The variable seeing affected our ability to achieve uniform sensitivities for all the target stars in the survey.  

To aid in the suppression of speckle noise, we observed a set of PSF stars in conjunction with 
about half of our targets.  The PSF stars were selected to have similar colors (i.e. spectral types), brightness, and airmasses as the targets so as to produce a speckle pattern as close as possible to the science targets. Finally, sets of known binary stars with high quality 
orbital solutions were also observed during the 2005 November and 2004 October runs 
to provide an accurate determination of the plate scale and image orientation. 
During the observation, each binary was placed in multiple positions over the 
field of view of the camera. In order to observe as many targets as possible while also reaching image
sensitivities necessary to detect brown dwarf companions, our observing goal was to complete all observations 
associated with each target within an half an hour.

\section{Data Reduction and Analysis}

\begin{figure}[ht]
\plottwo{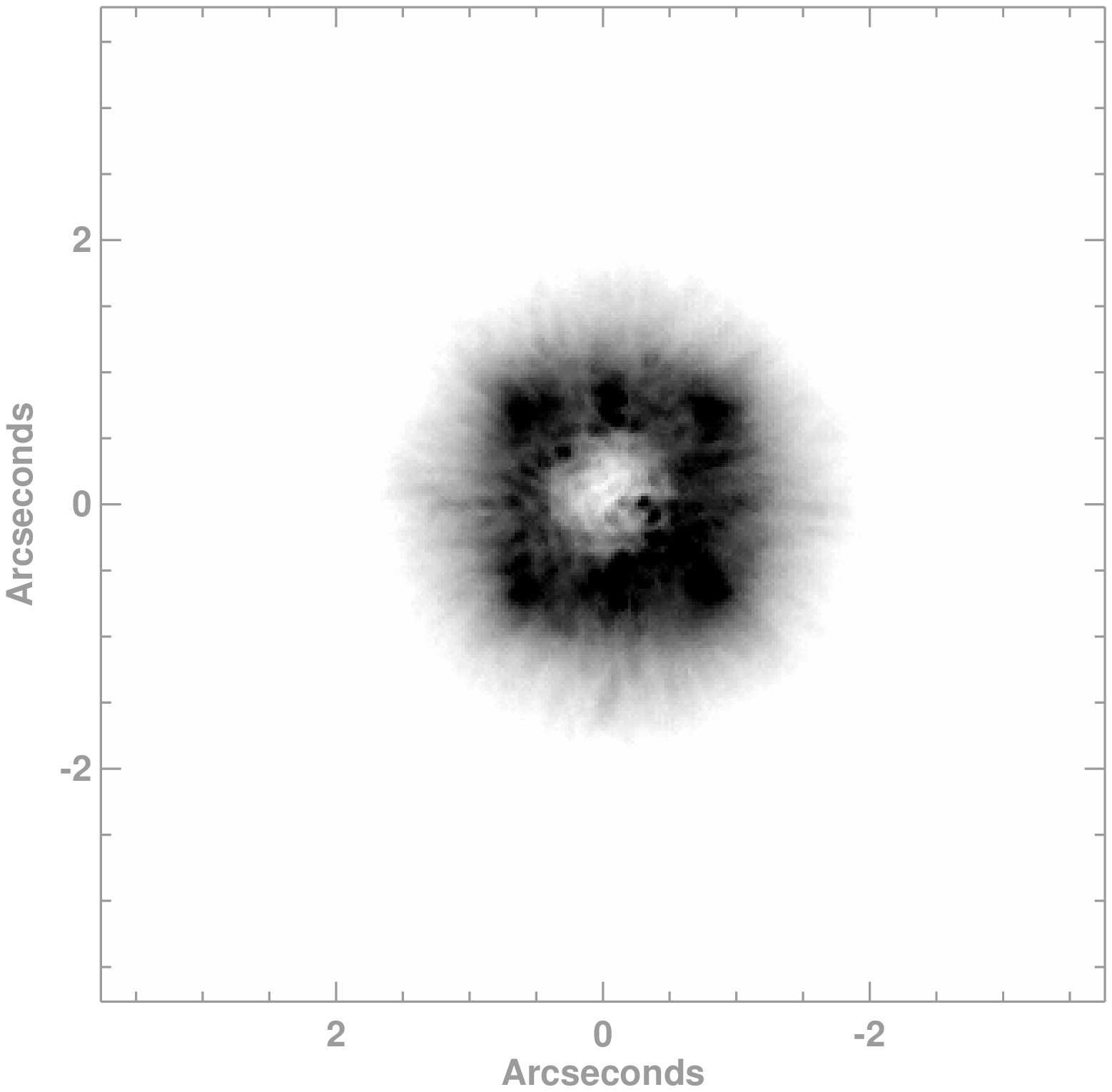}{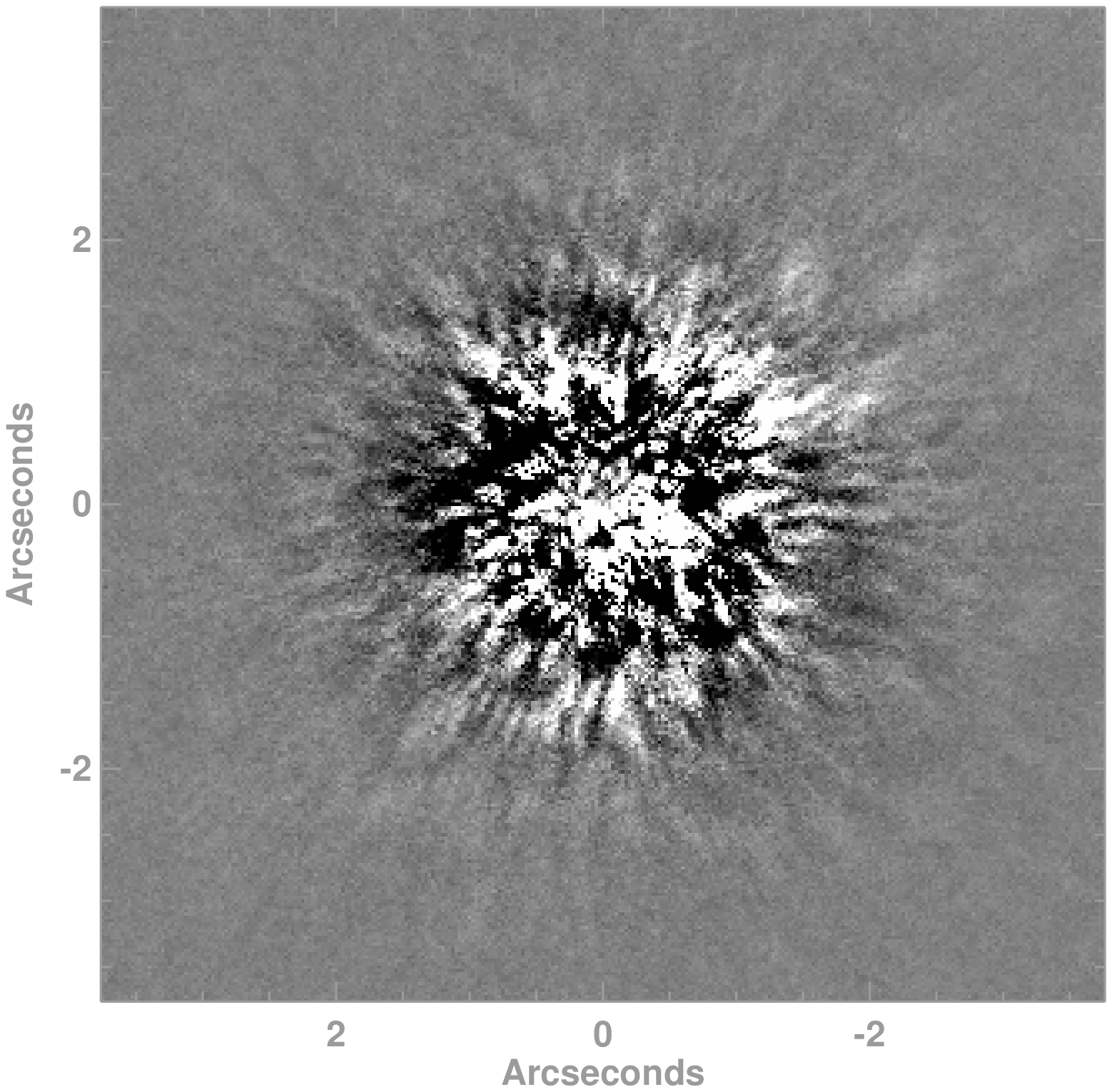} 
\figcaption{{\it Left} - Image of GL 876 prior to PSF subtraction. {\it Right} - The difference image between GL 876 and HD 216789 showing the reduced flux
in the halo of the PSF as a result of the subtraction.  North is
up and East is to the left in all the images.  \label{diffim}}
\end{figure}

The individual AO images were sky-subtracted, flat-fielded and corrected for bad-pixels with the
final image created from the median of the stack of reduced individual images. For those images with
additional stars visible in the field of view, we used them to determine the offset of each image in the stack with respect to
a reference image and shifted them accordingly prior to taking the median of the stack. 
This helped mitigate smearing from movement of the field during the observations due to 
telescope flexure or small losses in telescope tracking. A subset of the target images
have corresponding observations of PSF stars with similar brightness and color. 
The relative positions of the target and PSF star were estimated using the Poisson spot present
in the middle of the coronagraphic spot. As with previous studies (Metchev et al.  2004; Tanner et al. 2007) we
are able to register the star positions to 0.7 pixels using the centroid position of the spot (Tanner et al. 2007). 
The scaling for the PSF is the multiplicative factor which minimizes the residuals remaining
after the subtraction. Comparisons of the
standard deviations of the speckle noise of the stars before and after PSF subtraction 
suggest that the subtraction reduced the noise within the halo by $\sim$25\%.
Figure~\ref{diffim} shows the difference image of GL 876 and its PSF star HD 216789.

The coronagraphic images were flux calibrated using the images 
taken with the primary star off-set from the coronagraph while accounting
for a difference in integration time and the well-defined neutral density
filter used for the off-spot images (Metchev et al. 2004). 
The magnitudes for both the primary stars in the off-spot images
and the companions in the coronagraphic images are 
estimated from aperture photometry with an aperture of 0.8$\arcsec$ and sky 
annulus of 1.-1.25$\arcsec$. The K$_s$ band magnitudes of the primaries were 
taken from the Two Micron All-Sky Survey (Cutri et al. 2006). The uncertainties for the
photometry were estimated from the errors given for the 2MASS magnitude and an 
assumption of a 5\% calibration error determined by comparing the photometry
of the companion to GJ 105a to published values (Golimowski et al. 2000). 

The images of the calibration binaries were reduced in the standard manner.  
For the 2004 data, we assume a plate scale of 25.11$\pm$0.04 mas pixel$^{-1}$ 
estimated from the known orbital solutions of three different binary stars (WDS 09006+4147, WDS 18055+0230, 
WDS 20467+1607) that were observed very close to our October observations using the 
same instrument (Oct 4-5, 2004, Metchev 2005, see their Table 4.1). For the 2005 data we estimate 
a plate scale of 25.21$\pm$0.36 mas pixel$^{-1}$ using the average and standard 
deviation of the measured pixel separation of one binary (WDS 09006+4147, Hartkopf et al. 2006) compared to its
predicted orbital separation in arcseconds. This binary, which was observed in November 2005, 
was placed in multiple positions across the field of view after correction for the known distortion in the camera.  

\begin{figure}[ht]
\epsscale{1.0}
\plottwo{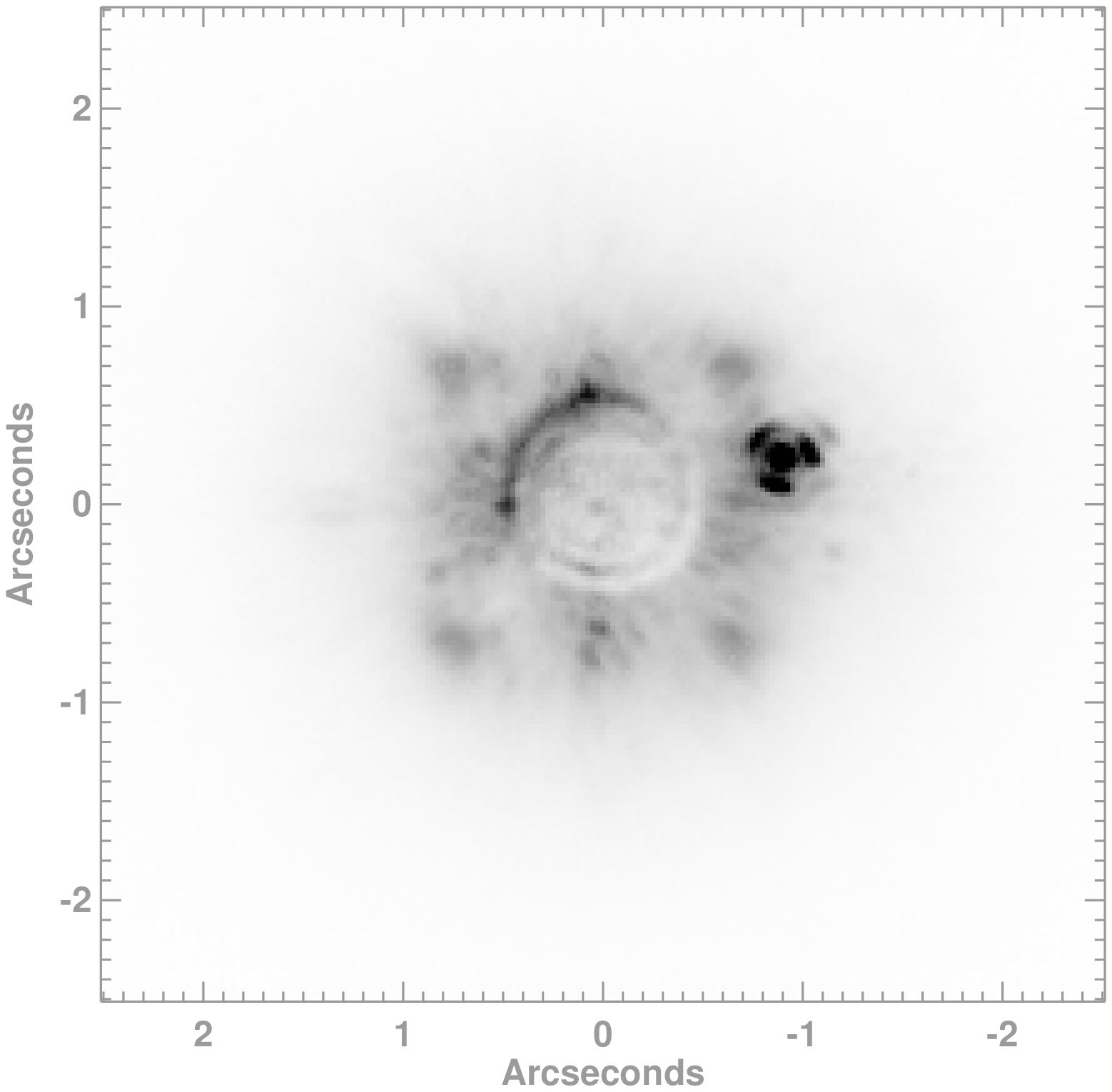}{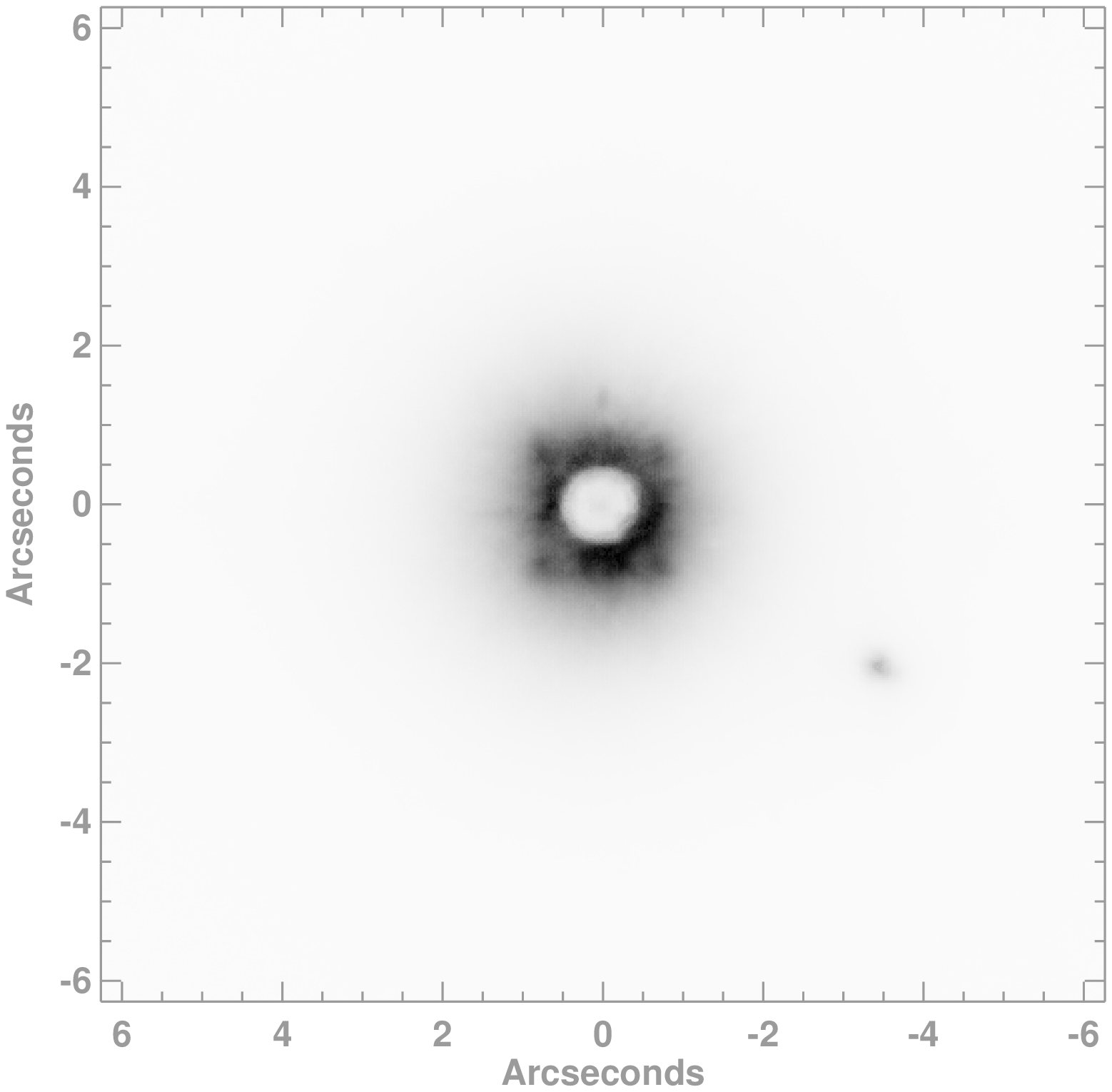}
\figcaption{ {\it Left-} Image of GL 454 and its common proper motion companion with a K$_s$ magnitude of 7.75 and a separation of 0.95$\arcsec$.  
{\it Right -}  Image of GL 1020 with a K$_s$=10.34 magnitude companion 4" away.  \label{finalim}}
\end{figure}

A thorough visual inspection of both the median-averaged, coronagraphic images
and the difference images between the targets and the PSF identified all stars in the 25$\arcsec$ field of view. 
Those target stars with no additional stars visible in the field-of-view  are listed in Table~\ref{nondet} while the companion candidates identified in these images are listed in Table~\ref{comptab} along with their distance from the target star, position angle and magnitude 
difference compared to the primary when available. To accurately determine the position of the star behind the coronagraphic spot, we utilized the
static waffle pattern which has a distinct set of four speckles framing the PSF (see Figure~\ref{finalim}). As determined in Tanner et al. (2007), 
by using the centroid position of each of the four spots in the waffle pattern, we can use the intersection
of the lines crisscrossing the pattern to determine the star's position to 2.3 pixels. This positional accuracy and the error in the plate scale
are then propagated when determining the uncertainties for the offset and position angle.

\subsection{Common Proper Motion Determination}
 
For thirteen of the stars with companion candidates, we collected second epoch observations at least a year later during AO observing runs dedicated for other projects. The additional
image allows us to determine whether the companion is bound to the star through their common proper motion. 
This positional accuracies given in Table~\ref{comptab} are sufficient to conclude whether the companion candidates are bound to their stars given the high proper motions  ($>$ 100 mas/yr ) of all the stars in the sample and the minimum of a full year between observations. In most
cases, simply blinking between the images shows the movement of the background star once the target star is held fixed. 
We determined the separation and position angles of the candidates in both images
and compared them to what would be expected given the star's proper motion. After this analysis, we find
two stars with confirmed common proper motion companions - GJ 454 and GJ 1020. The companions to these stars
are new detections having not been mentioned in previous publications. 

\begin{figure}[ht]
\plottwo{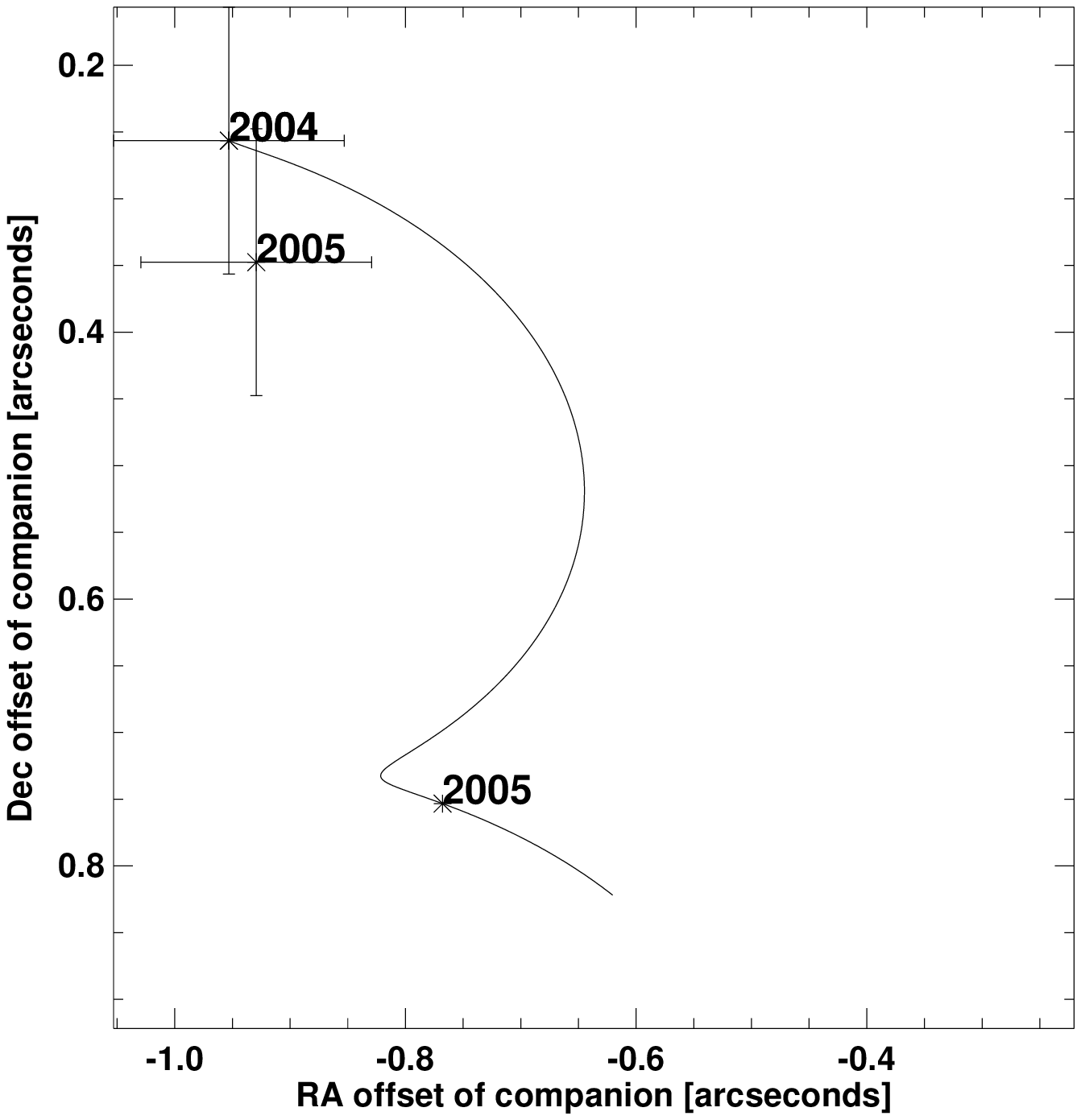}{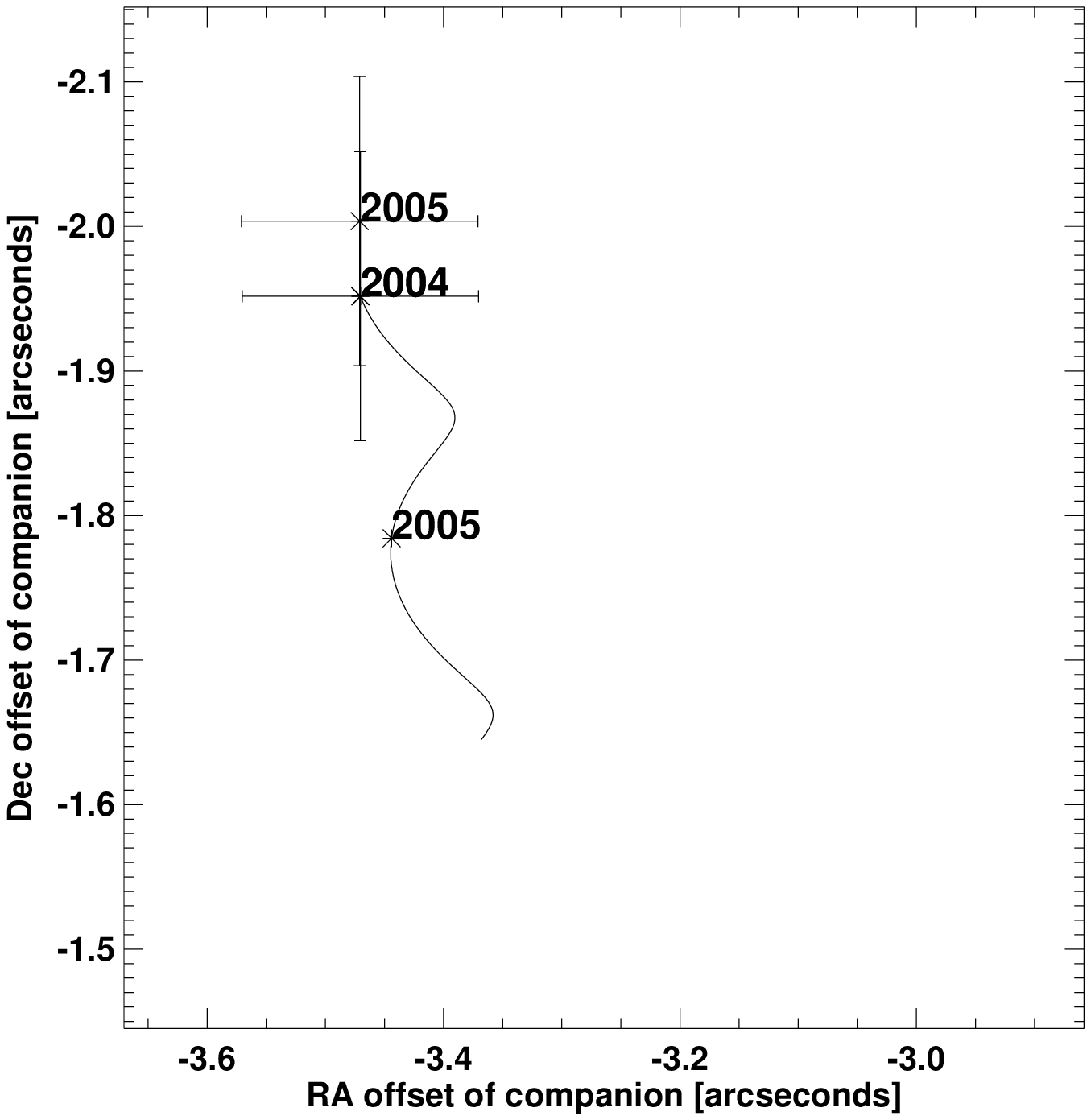} 
\figcaption{ Plots of the offsets in RA and Dec for the companion candidates around
GL 454 and GL 1020. The squiggly line denotes the expected change in the
offset if the companion were a background object. The labels (i.e. 2004, 2005) indicate the 
epochs of the observed offsets and the expected offsets of a background
object at the epochs of the observations. The fact that the observed 
offsets are consistent with each other within the uncertainties 
suggests these two companions are bound to the star. \label{pmplots}}
\end{figure}

Figure~\ref{pmplots} plots the offset in RA and Dec between GL 454 and its companion 1$\arcsec$ away and Table~\ref{compastron} lists the astrometry. The solid line 
denotes the expected motion of the companion if it were a stationary 
background object. The fact that the offset of the companion star estimated for
two different epochs does not change significantly given the associated errors (10-30 mas), confirms this object as a bona
fide companion. Additional observations of GJ 454 at 1.25 $\mu$m where made with the same
instrument during the June 2004 run and were reduced in the same way as the K$_S$ data. 
Given the K$_s$ and J magnitudes of the companion (K$_s$=7.75 and J=9.07) and
the distance to the star (12.91 pc), it is most likely an M3 dwarf (Baraffe \& Chabrier 1996). 
Despite its proximity to GJ 454, its relative brightness, $\Delta$K$_S >$ 6, it should not pose a problem for SIM Lite.
Figure~\ref{pmplots} also shows the astrometry for the companion to GJ 1020, which was originally observed as a PSF star to GJ 10.  This
 confirmed proper motion companion has  a K$_s$ magnitude of 10.4. With an absolute K magnitude of $>$ 10, this companion is most likely a
 low-mass star. 
 
For some of the targets with no second epoch observations in our survey, we 
utilized images from the HST\footnote{http://archive.stsci.edu/hst/}, ESO\footnote{http://archive.eso.org/cms/user-portal} and Gemini\footnote{http://www1.cadc-ccda.hia-iha.nrc-cnrc.gc.ca/gsa/}  archives. 
We were able to use these archive data sets to confirm that companion candidates around
GJ 726, GJ 722 and GJ 892 are background stars.  There are  17  companion candidates that remain unconfirmed (see Table~\ref{comptab}). 

\section{Results}

Out of 84 stars, we found two unpublished common proper motion companions to GJ 454 and GJ 1020, neither of which will 
cause problems for SIM Lite observations since it is much fainter than the target star. None of the companions have absolute magnitudes or colors
consistent with brown dwarfs. This is not unexpected given the 
observed paucity of brown dwarf companions to solar mass main sequence stars (McCarthy \& Zuckerman 2004; Butler et al. 2006).

\subsection{Image Sensitivities}

To estimate the sensitivities of all of our target fields as a function of distance from the star, we employ ``PSF planting''
in which a PSF corresponding to an object of known brightness is inserted into the 
image. The PSF extracted from the off-spot, flux-calibration image of each target is sky subtracted, 
normalized, multiplied by an array of contrast values ($\Delta K_s$=7.7-15.1 mags) 
and placed at a 
range (0-10$\arcsec$) of distances from the target at random position
angles. We completed 10$^4$ iterations of the PSF planting algorithm 
to fill out the parameter space of contrast and distance from the primary star. We make sure that the 
same number of planted PSFs occur in each radius bin. 
To determine whether the planted star is detected, the image is cross-correlated 
with a flux normalized PSF. For each distance bin we 
estimate the minimum PSF intensity for which at least 90\% of the fake sources had a 
correlation value of 0.75 or higher. This correlation value was determined by visually inspecting an
image with inserted scaled PSFs. Previous studies have shown that the eye
is often the best natural detector when determining the presence of a faint 
star in an image (Metchev et al. 2005).

\begin{figure}[ht]
\plotone{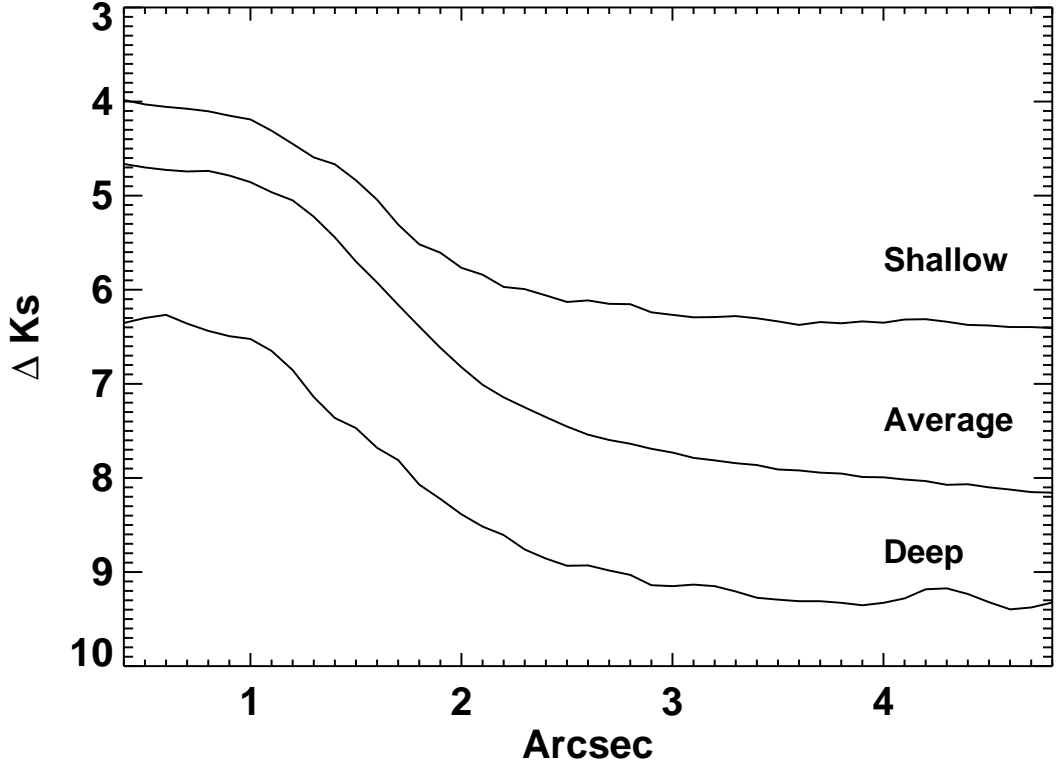} 
\figcaption{ Plot of the differential magnitude detectable in the PHARO images
as a function of distance from the star in arcseconds. The three lines represent the averages of three levels of sensitivity which corresponds to 
values at 3$\arcsec$ of $\Delta$K$_s>$6, 5$<$$\Delta$K$_s$$<$4, and $\Delta$K$_s$$<$4 magnitudes, respectively. The sensitivity of any image depends both on the integration time of the exposure and the seeing conditions.  \label{senplot}}
\end{figure}

The intensities are converted into magnitudes using the flux 
calibration from the off-spot image and the 2MASS K$_s$ magnitude of the star. 
Figure~\ref{senplot} plots the largest K$_s$ magnitude difference 
between the target star and planted PSF as a function 
of distance from the star for all targets with calibration
data. 

\begin{figure}[ht]
\plotone{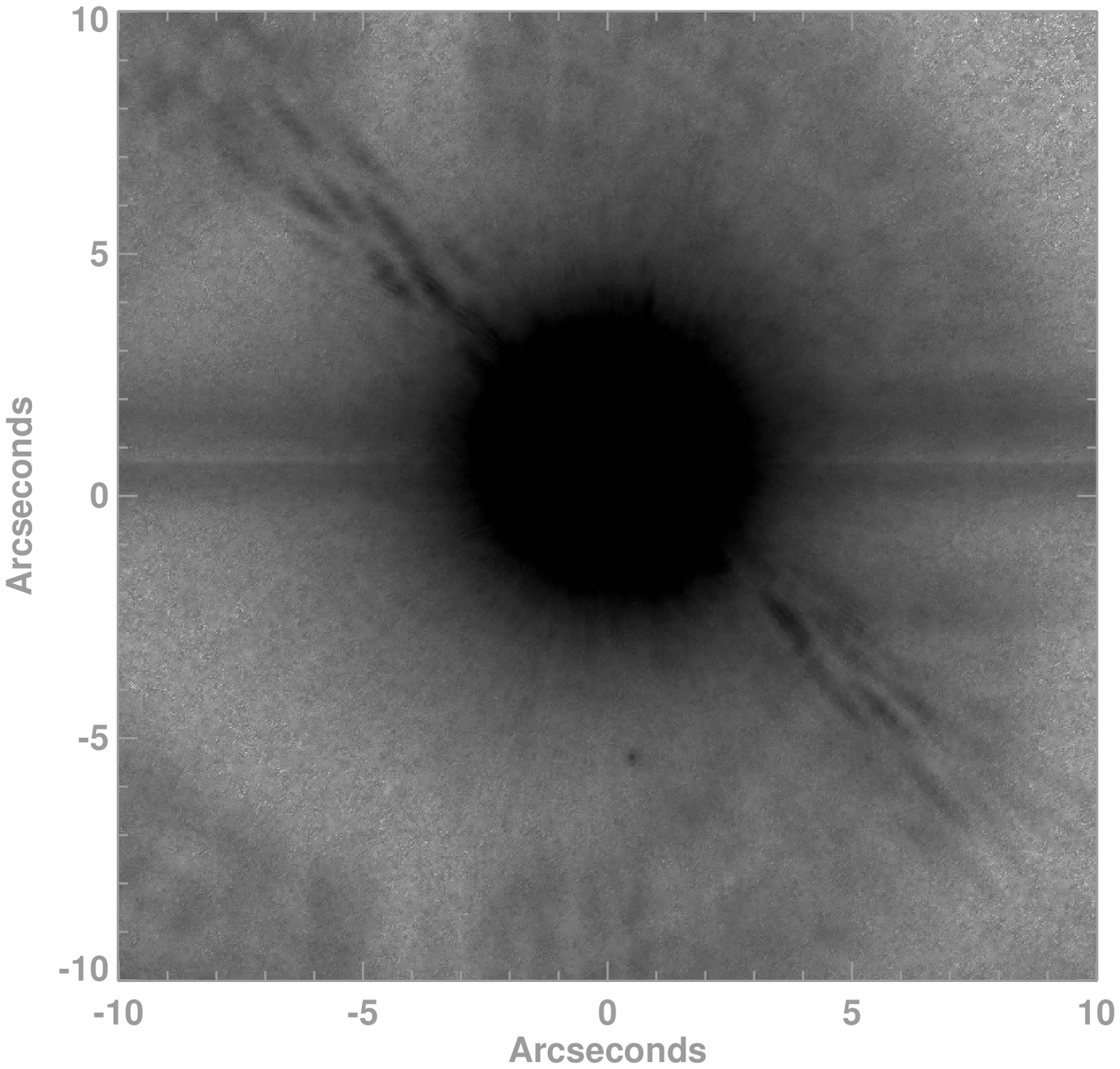} 
\figcaption{ Image of GL 15 A using a histogram stretch to emphasize the scattered 
light from impurities in the optical path. These reflections, which are
worse for bright stars, reduce contrast sensitivities by 15-20\%. \label{gl15stretch}}
\end{figure}

Table~\ref{sentab} lists the values of the faintest detectable K$_s$ magnitudes 
at 0.5, 1, 2, and 5  arcseconds. We were
able to detect sources with a magnitude contrast 
of  $\Delta K_S$$\sim$4-6.5 mag at 0.5$\arcsec$, $\Delta K_S$$\sim$5.5-8 mag at 2$\arcsec$ and $\Delta K_S$$\sim$6.5-9.5 at 5$\arcsec$
with $\sim$90$\%$ completeness. 
  The range of image contrasts is due primarily to variations in seeing conditions 
throughout the night. Unfortunately, the brightest objects
suffer from two additional components of degradation in image sensitivity: 1) a smudge in the 
optics which creates a long streak across the spot at a position angle of 45 degrees, and 2) a 
thick, off-set ring due to a reflection off the array, entrance windows, or
the coronographic slide which is illuminating the lyot stop
(see Figure~\ref{gl15stretch}). These image artifacts reduce the sensitivity in these areas of the image by 15-20\%.

\subsection{Mass Sensitivities}

\begin{figure}[ht]
\epsscale{0.9}
\plotone{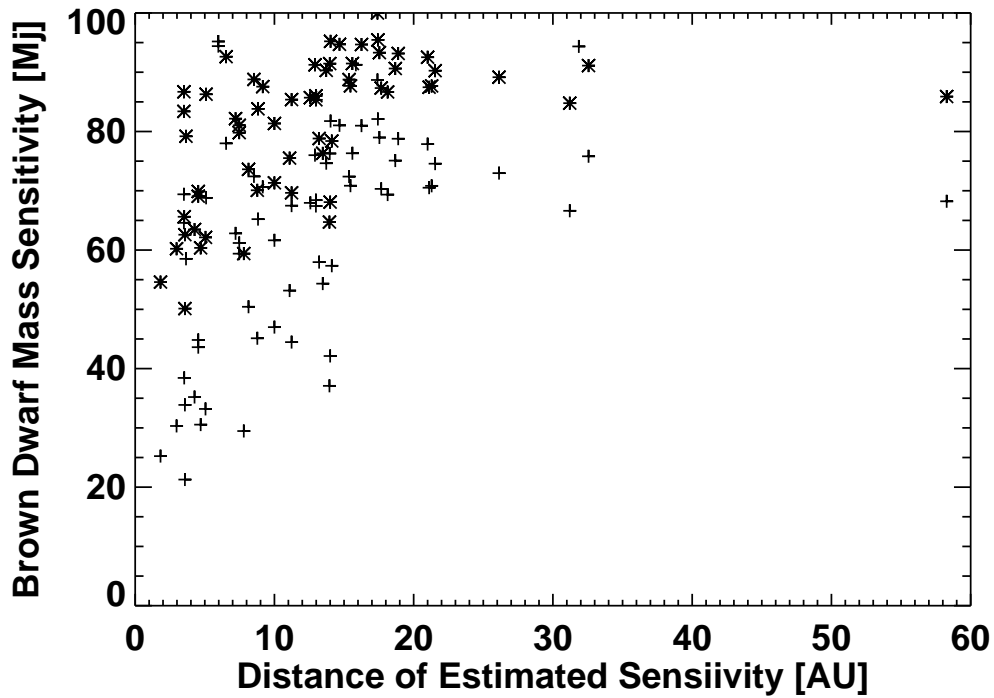} 
\figcaption{Minimum companion mass limits (Burrows et al. 1997) estimated from PSF planting as a function of projected physical distance from the star assuming ages of both 1 Gyr (+) and 5 Gyr (*).  We are sensitive to brown dwarf mass companions (M $<$75 M$_J$) for many of the stars in the survey.   \label{senvsdistance}}
\end{figure}

By assuming a set of values relating the absolute magnitude of any brown dwarf companion to its mass
and age based on theoretical evolution models (Burrows et al. 1997; Baraffe et al. 2003), we can estimate an lower limit to the
mass sensitivity around those target stars with flux calibrations. Figure~\ref{senvsdistance} plots the lower limits 
of the mass sensitivity at a separation of 1$\arcsec$ from each star as a function of separation in AU based on its Hipparcos distance. 
The upper limits assume both 1 Gyr (crosses) and 5 Gyr (asteriks) ages for the primary star. 
This plot shows that most of our observations are reaching into the brown dwarf regime ($\sim$ 75 M$_J$) at separations of 1" and beyond. 

\section{Discussion}

\subsection{How does this survey compare with previous work?}

There have been a few high contrast surveys consisting of similar samples of nearby, mature stars. Of those, the most notable include
McCarthy \& Zuckerman (2004), a large non-AO coronagraphic survey of a couple hundred nearby, mature stars from which the ``brown dwarf desert'' was first posited at large ($>$100 AU) separations. 
Soon after this survey suggested that the percentage of mature stars with wide brown dwarf companions was comparable to the numbers 
determined from the RV surveys, $\sim$1\%, additional coronagraphic AO surveys of similar targets suggested that this value is larger (3-10\%) 
depending on the age of the targets and assuming non-circular orbits (Carson et al. 2006; Lowrance et al. 2005; Metchev et al. 2009). Based our mass sensitivities and level of completeness, at a separation of one arcsecond ($>$5-60 AU), we can place a 3 $\sigma$ upper limit of 17\% on brown dwarf companions with masses down to 40 M$_J$ at an age of 1 Gyr.  

This survey presented here increases the sample of nearby, mature ($>$1 Gyr) stars with high contrast AO observations by over 50\% when considering
the ``field'' surveys listed in the table from Metchev et al. (2009). However, since the survey was
originally designed as a precursor program to look for stellar companions to SIM Lite targets, our sample is not complete in a way that
would allow us to provide improved limits on the brown dwarf companion fraction for the separations probed. However, these observations will be useful to future imaging surveys like GPI and SPHERE which are expected to be sensitive to planetary-mass objects as well as brown dwarfs. 

 \subsection{The Value of publishing non-detections}
 
With projects like GPI, SPHERE and SIM Lite being developed to focus almost primarily on the
detection of exoplanets, it has become necessary to compile samples of targets which are going to yield the largest number
of detections given the strengths and weaknesses of the instruments. Therefore, many years before these projects see first light,
much effort is devoted to target selection, verification and precursor observations (i.e. Tanner et al. 2007; Carson et al. 2007).  As a part of these precursor programs
and as separate high resolution studies focused on detecting brown dwarf and planetary mass objects at wide separations, a large sample
of stars have been observed with adaptive optics, coronagraphic instruments. Unfortunately, in most cases, only those stars with companion candidates
or confirmed companions are published. As a result, there are numerous unpublished observations of potential targets resulting in repetitive observations
of the same star in multiple studies. While these studies most likely have different inner working angles and overall sensitivities, there is value in 
having a public archive of previous observations to either weed out unknown, close stellar binaries or to facilitate the identification of common proper motion.
In fact, it was archival images of HR 8799 from a previously unpublished AO survey of nearby stars that was used to detect the orbital motion
of the directly imaged planets in this system (Marois et al. 2009). In addition, the absence of companions up to a given 
mass at a given separation for a large sample of stars can provide just as much information on planetary formation and dynamics 
as their detection. With this in mind, all of the AO images collected for this survey will be donated to the NStED\footnote{http://nsted.ipac.caltech.edu}  database and we encourage others to do the same. 

\section{Conclusions} 

We have completed a high contrast imaging survey of a sample of stars slated to be
potential targets for the SIM Lite astrometric space telescope. While many of our observations
were sensitive enough to detect brown dwarf companions at separations of $>$1$\arcsec$, our survey
found two unpublished confirmed common proper motion stellar companions around GJ 454 and GJ 1020. 
This survey will serve as a resource for both SIM Lite and direct imaging surveys such
as GPI and SPHERE as all the images will be given to the NStED database. These images can be used to vet future targets
and aid in common proper motion determinations. Additional SIM-Lite targets will be observed with high contrast imaging if they have 
not all ready been observed with other programs. 

\begin{acknowledgements}

We would like to thank our anonymous referee for their valuable insights into how to improve our
manuscript. We'd also like to thank Shri Kulkarni and Mike Shao for invaluable discussions regarding the selection of the EPIcS targets. 
Based on observations obtained at the Hale Telescope, Palomar Observatory,  
as part of a continuing collaboration between the California Institute of 
Technology, NASA/JPL, and Cornell University. The research described in this 
publication was carried out at the Jet Propulsion Laboratory,
California Institute of Technology, under a contract with the National 
Aeronautics and Space Administration. This publication makes use of data products 
from the Two Micron All Sky Survey, which is a joint project of the University of 
Massachusetts and the Infrared Processing and Analysis Center/California Institute 
of Technology, funded by the National Aeronautics and Space Administration and 
the National Science Foundation.
\end{acknowledgements}

\begin{deluxetable}{lrrrrcccccc}	
\footnotesize																				
\tablecaption{Palomar  AO Sample\label{sample}}																			
\tablehead{																					
\colhead{Target}	&	\colhead{RA}	&	\colhead{Dec}	&	\colhead{$\mu_\alpha$}	&	\colhead{$\mu_\delta$}	&	\colhead{V}	&	\colhead{Ks}	&	\colhead{SpTy}	&	\colhead{Distance}	&	\colhead{Observing Dates}	&	\colhead{t$_{int}$}	\nl
\colhead{}	&	\colhead{}	&		&	\colhead{mas/yr}	&	\colhead{mas/yr}	&	\colhead{}	&	\colhead{}	&	\colhead{}	&	\colhead{pc}	&		&	\colhead{sec}	
}																					
\startdata	
GJ 10	&	00 11 15.86	&	-15 28 04.72	&	-84	&	-269	&	4.89	&	3.82	&	 F8V	&	18.89	&	Aug 2004; Nov 2005	&	75; 400	\nl
GJ 15 A    	&	00 18 22.89	&	+44 01 22.63	&	2889	&	410	&	8.07	&	4.02	&	M2V	&	3.57	&	Oct 2004; Nov 2005	&	300; 500	\nl
GJ 15 B    	&	00 18 25.87	&	+44 01 38.44	&	2912	&	351	&	11.04	&	5.95	&	M3.5	&	3.57	&	Oct 2004; Nov 2005	&	1200; 500	\nl
GJ 1020 	&	00 45 28.69	&	-12 52 50.92	&	-32	&	-206	&	6.15	&	4.62	&	G0V	&	31.86	&	Dec 2004; Nov 2005	&	400; 200	\nl
GJ 34 B    	&	00 49 05.17	&	+57 49 03.77	&	1105	&	-493	&	7.51	&	3.88	&	K7V	&	5.95	&	Aug 2004	&	105	\nl
GJ 34 A    	&	00 49 06.29	&	+57 48 54.67	&	1087	&	-560	&	3.45	&	1.99	&	G0V	&	5.95	&	Aug 2004	&	150	\nl
GJ 37	&	00 50 07.59	&	-10 38 39.57	&	-225	&	-228	&	5.19	&	4.02	&	F7IV-V	&	15.46	&	Dec 31 2004; Nov 2005	&	300; 400	\nl
GJ 61  A    	&	01 36 47.84	&	+41 24 19.65	&	-173	&	-381	&	4.09	&	2.86	&	F8V	&	13.47	&	Aug 2004; Nov 2005	&	45; 100	\nl
GJ 68	&	01 42 29.76	&	+20 16 06.62	&	-302	&	-677	&	5.20	&	3.29	&	K1V	&	7.47	&	Aug 2004; Dec 2005	&	45; 350	\nl
GJ 71	&	01 44 04.08	&	-15 56 14.93	&	-1722	&	854	&	3.50	&	1.79	&	G8V	&	3.65	&	Aug 2004	&	4.5	\nl
GJ 72	&	01 44 55.82	&	+20 04 59.34	&	-45	&	-105	&	6.29	&  4.60	&	G5IV	&	32.56	&	Aug 2004; Dec 2005	&	150; 140	\nl
GJ 79	&	01 52 49.17	&	-22 26 05.48	&	844	&	-1	&	8.88	& 5.18	&	K9Vk	&	11.09	&	Aug 2004	&	150	\nl
GJ 105 A   	&	02 36 04.89	&	+06 53 12.73	&	1806	&	1442	&	5.82	&	3.48	&	K3V	&	7.21	&	Aug 2004	&	45	\nl
GJ 3175	&	02 40 12.42	&	-09 27 10.35	&	-138	&	-79	&	5.80	&	4.53	&	F6V	&	21.54	&	Dec 2004	&	790	\nl
GJ 107 B   	&	02 44 10.26	&	+49 13 54.06	&	336	&	-84	&	9.87	&	5.87	&	M1.5	&	...	&	Oct 2004	&	1800	\nl
GJ 107 A   	&	02 44 11.99	&	+49 13 42.41	&	334	&	-90	&	4.12	&	2.70	&	F7V	&	11.23	&	Aug 2004; Oct 2004	&	100; 1800	\nl
GJ 111	&	02 45 06.19	&	-18 34 21.23	&	331	&	36	&	4.50	&	3.25	&	F6V	&	13.97	&	Aug 2004; Oct 2004	&	40; 250	\nl
GJ 115 A	&	02 50 41.42	&	-44 04 52.69	&	-24	&	-272	&	8.19	&	6.70	&	F8V	&	58.28	&	Aug 2004	&	75	\nl
GJ 137	&	03 19 21.70	&	+03 22 12.71	&	269	&	94	&	4.83	&	2.96	&	G5Vv	&	9.16	&	Aug 2004	&	45	\nl
GJ 147	&	03 36 52.38	&	+00 24 05.98	&	-233	&	-482	&	4.28	&	2.84	&	F9IV-V	&	13.72	&	Aug 2004	&	63	\nl
GJ 204	&	05 28 26.10	&	-03 29 58.40	&	-307	&	-797	&	7.64	&	4.88	&	K5V	&	12.98	&	Dec 2004	&	2400	\nl
GJ 9207	&	06 16 26.62	&	+12 16 19.79	&	83	&	186	&	5.04	&	4.24	&	F5IV-V	&	19.61	&	Nov 2005	&	3000	\nl
GJ 250 B	&	06 52 18.07	&	-05 11 25.6	&	-541	&	0	&	10.05	&	5.72	&	M2	&	...	&	Nov 2005	&	600	\nl
GJ 302 	&	08 18 23.95	&	-12 37 55.82	&	279	&	-989	&	5.95	&	4.17	&	K0V	&	12.58	&	Dec 2004	&	900	\nl
GJ 303 	&	08 20 03.86	&	+27 13 03.75	&	-18	&	-376	&	5.10	&	3.87	&	F6V	&	18.13	&	Nov 2005	&	400	\nl
GJ 338 A	&	09 14 22.79	&	+52 41 11.85	&	-1533	&	-563	&	7.64	&	3.99	&	M0V	&	6.19	&	June 2004	&	5.4	\nl
GJ 338 B	&	09 14 24.70	&	+52 41 10.95	&	-1551	&	-656	&	7.74	&	4.14	&	M0V	&	6.27	&	June 2004	&	5.4	\nl
GJ 382	&	10 12 17.67	&	-03 44 44.38	&	-153	&	-243	&	9.26	&	5.02	&	M1.5	&	7.81	&	June 2004	&	300	\nl
GJ 394	&	10 30 25.31	&	+55 59 56.83	&	-181	&	-34	&	8.76	&	5.36	&	K7V	&	10.99	&	June 2004	&	50	\nl
GJ 423.1	&	11 18 22.01	&	-05 04 02.29	&	795	&	-151	&	7.31	&	5.46	&	G8V	&	21.99	&	June 2004	&	50	\nl
GJ 447	&	11 47 44.40	&	+00 48 16.43	&	606	&	-1219	&	11.08	&	5.65	&	M4	&	3.34	&	June 2004	&	110	\nl
GJ 448	&	11 49 03.58	&	+14 34 19.42	&	-499	&	-114	&	2.14	&	1.88	&	A3V	&	11.09	&	June 2004	&	360	\nl
GJ 454	&	12 00 44.45	&	-10 26 45.65	&	142	&	-483	&	5.54	&	4.03	&	K0IV	&	12.91	&	June 2004; July 2005	&	3; 40	\nl
GJ 506	&	13 18 24.31	&	-18 18 40.31	&	-1070	&	-1064	&	4.74	&	2.96	&	G5V	&	8.53	&	June 2004; July 2005	&	5.4; 60	\nl
GJ 526	&	13 45 43.78	&	+14 53 29.47	&	1778	&	-1455	&	8.46	&	4.41	&	M2V	&	5.43	&	June 2004	&	10	\nl
GJ 527 A	&	13 47 15.74	&	+17 27 24.86	&	-480	&	54	&	4.50	&	3.51	&	F6IV	&	15.60	&	July 2005	&	60	\nl
GJ 555	&	14 34 16.81	&	-12 31 10.40	&	-358	&	595	&	11.35	&	5.94	&	M4	&	6.12	&	June 2004	&	270	\nl
GJ 557	&	14 34 40.82	&	+29 44 42.47	&	188	&	133	&	4.46	&	3.34	&	F2V	&	15.47	&	July 2005	&	90	\nl
GJ 9491	&	14 43 03.62	&	-05 39 29.54	&	104	&	-320	&	3.90	&	3.04	&	F2V 	&	18.68	&	Aug 2004	&	45	\nl
GJ 570 A	&	14 57 28.00	&	-21 24 55.71	&	1034	&	-1725	&	5.74	&	3.05	&	K4V	&	5.91	&	June 2004	&	480	\nl
GJ 581	&	15 19 26.83	&	-07 43 20.21	&	-1225	&	-100	&	10.57	&	...	&	M3	&	6.27	&	June 2004	&	480	\nl
GJ 598	&	15 46 26.61	&	+07 21 11.06	&	-226	&	-69	&	4.43	&	2.99	&	G0V	&	11.75	&	June 2004	&	600	\nl
GJ 602	&	15 52 40.54	&	+42 27 05.47	&	439	&	630	&	4.62	&	2.58	&	F8Ve...	&	15.85	&	July 2005	&	60	\nl
GJ 606.2	&	16 01 02.66	&	+33 18 12.63	&	-197	&	-773	&	5.40	&	3.86	&	G0Va	&	17.43	&	July 2005	&	90	\nl
GJ 616	&	16 15 37.27	&	-08 22 09.99	&	232	&	-496	&	5.50	&	4.19	&	G2Va	&	14.03	&	June 2004; Aug 2004	&	960	\nl
GJ 628	&	16 30 18.06	&	-12 39 45.34	&	-94	&	-1185	&	10.12	&	...	&	M3.5	&	4.26	&	June 2004; July 2004	&	10; 450	\nl
GJ 629.1 	&	16 32 57.88	&	-12 35 30.23	&	-313	&	-226	&	10.61	&	7.25	&	M0	&	31.21	&	June 2004	&	450	\nl
GJ 663 A   	&	17 15 20.85	&	-26 36 09.04	&	-488	&	-1156	&	5.29	&	...	&	K0V	&	5.46	&	June 2004	&	7.5\nl
GJ 663 B   	&	17 15 20.98	&	-26 36 10.18	&	-473	&	-1143	&	5.33	&	...	&	K1.5V	&	5.99	&	June 2004	&	7.5	\nl
GJ 664	&	17 16 13.36	&	-26 32 46.13	&	-480	&	-1123	&	6.34	&	...	&	K5V	&	5.97	&	June 2004	&	10	\nl
GJ 670 AB	&	17 21 00.37	&	-21 06 46.56	&	262	&	-205	&	4.39	&	3.07	&	F2V	&	17.40	&	June 2004	&	4	\nl
GJ 678 A	&	17 30 23.52	&	-01 03 54.6	&	-116	&	-170	&	6.00	&	...	&	G8IV-V	&	...	&	June 2004	&	90	\nl
GJ 687	&	17 36 25.90	&	+68 20 20.91	&	-320	&	-1270	&	9.15	&	4.55	&	M3.5V	&	4.53	&	June 2004; Nov 2005	&	12; 150	\nl
GJ 692	&	17 43 25.79	&	-21 40 59.50	&	-98	&	-45	&	4.87	&	3.88	&	F5V	&	17.54	&	June 2004	&	5.4	\nl
GJ 699	&	17 57 48.50	&	+04 41 36.25	&	-799	&	10338	&	9.54	&	4.52	&	M4Ve	&	1.82	&	June 2004; Nov 2005	&	12; 120	\nl
GJ 701	&	18 05 07.58	&	-03 01 52.75	&	570	&	-333	&	9.37	&	...	&	M1	&	7.80	&	June 2004; Aug 2004	&	5.4	\nl
GJ 702 A	&	18 05 27.37	&	+02 29 59.32	&	276	&	-1092	&	4.20	&	1.79	&	K0V	&	5.09	&	June 2004; Aug 2004	&	90	\nl
GJ 702 B	&	18 05 27.42	&	+02 29 56.42	&	442	&	-1253	&	6.00	&	...	&	K4V	&	5.09	&	June 2004; Aug 2004	&	6; 45	\nl
GJ 716	&	18 31 18.96	&	-18 54 31.72	&	-140	&	-195	&	6.82	&	4.70	&	K2V	&	13.21	&	June 2004; Aug 2004	&	10	\nl
GJ 722	&	18 38 53.40	&	-21 03 06.74	&	-75	&	-152	&	5.87	&	4.23	&	G6V	&	12.98	&	June 2004; Aug 2004	&	10	\nl
GJ 725 A	&	18 42 46.69	&	+59 37 49.43	&	-1327	&	1802	&	8.91	&	4.43	&	M3V	&	3.57	&	June 2004; Aug 2004	&	12; 150	\nl
GJ 725 B	&	18 42 46.90	&	+59 37 36.65	&	-1393	&	1846	&	9.69	&	5.00	&	M3.5	&	3.52	&	June 2004; Aug 2004	&	12; 150	\nl
GJ 726	&	18 47 27.25	&	-03 38 23.39	&	-133	&	-273	&	8.81	&	5.58	&	K5	&	14.12	&	June 2004	&	180	\nl
GJ 729	&	18 49 49.36	&	-23 50 10.44	&	638	&	-192	&	10.95	&	5.37	&	M3.5	&	2.97	&	June 2004; Aug 2004	&	10	\nl
GJ 768	&	19 50 47.00	&	+08 52 05.96	&	537	&	386	&	0.77	&	0.10	&	A7V	&	5.14	&	June 2004	&	2	\nl
GJ 779	&	20 04 06.22	&	+17 04 12.62	&	-394	&	-406	&	5.80	&	4.39	&	G0V	&	17.67	&	June 2004; Aug 2004	&	4; 42	\nl
GJ 785	&	20 15 17.39	&	-27 01 58.72	&	1241	&	-181	&	5.73	&	3.50	&	K2	&	8.82	&	July 2004	&	150	\nl
GJ 789	&	20 22 52.37	&	+14 33 03.95	&	79	&	-7	&	6.17	&	4.90	&	F8V	&	26.13	&	Aug 2004	&	100	\nl
GJ 796	&	20 40 11.76	&	-23 46 25.92	&	501	&	461	&	6.37	&	4.60	&	G8V	&	14.65	&	June 2004	&	130	\nl
GJ 805	&	20 46 05.73	&	-25 16 15.23	&	-51	&	-157	&	4.15	&	3.09	&	F5V	&	14.67	&	July 2004	&	150	\nl
GJ 811	&	20 56 47.33	&	-26 17 46.96	&	95	&	-65	&	5.70	&	4.48	&	F6V	&	21.00	&	July 2004	&	150	\nl
\enddata
\end{deluxetable}	

\begin{deluxetable}{lrrrrcccccc}	
\footnotesize																				
\tablecaption{Palomar  AO Sample - cont.\label{sample}}																			
\tablehead{																					
\colhead{Target}	&	\colhead{RA}	&	\colhead{Dec}	&	\colhead{$\mu_\alpha$}	&	\colhead{$\mu_\delta$}	&	\colhead{V}	&	\colhead{Ks}	&	\colhead{SpTy}	&	\colhead{Distance}	&	\colhead{Observing Dates}	&	\colhead{t$_{int}$}	\nl
\colhead{}	&	\colhead{}	&		&	\colhead{mas/yr}	&	\colhead{mas/yr}	&	\colhead{}	&	\colhead{}	&	\colhead{}	&	\colhead{pc}	&		&	\colhead{sec}	
}																					
\startdata
GJ 820 A	&	21 06 53.94	&	+38 44 57.90	&	4157	&	3259	&	5.21	&	2.25	&	K5V	&	3.48	&	June 2004; Nov 2005	&	5.4	\nl
GJ 820 B	&	21 06 55.26	&	+38 44 31.40	&	4109	&	3144	&	6.03	&	2.54	&	K7V	&	3.50	&	June 2004; Nov 2005	&	5.4	\nl
GJ 821	&	21 09 17.42	&	-13 18 09.02	&	710	&	-1995	&	10.87	&	...	&	M1	&	12.15	&	June 2004; Nov 2005	&	270	\nl
GJ 848.4 A	&	22 09 29.87	&	-07 32 55.16	&	85	&	-450	&	6.63	&	4.89	&	G0V	&	21.29	&	June 2004	&	270	\nl
GJ 849	&	22 09 40.35	&	-04 38 26.62	&	1135	&	-20	&	10.42	&	5.59	&	M3.5	&	8.77	&	Aug 2004	&	500	\nl
GJ 872 A	&	22 46 41.58	&	+12 10 22.40	&	233	&	-492	&	4.20	&	2.96	&	F7V	&	16.25	&	June 2005	&	60	\nl
GJ 873	&	22 46 49.73	&	+44 20 02.37	&	-705	&	-459	&	10.09	&	5.30	&	M3.5	&	5.05	&	June 2004; Nov 2005	&	10; 600	\nl
GJ 875	&	22 50 19.43	&	-07 05 24.39	&	-103	&	103	&	9.97	&	6.10	&	K7	&	14.00	&	Oct 2004	&	1800	\nl
GJ 876	&	22 53 16.73	&	-14 15 49.32	&	960	&	-676	&	10.17	&	...	&	M4	&	4.70	&	July 2004	&	650	\nl
GJ 882	&	22 57 27.98	&	+20 46 07.80	&	208	&	61	&	5.49	&	3.91	&	G5V	&	15.36	&	June 2005	&	180	\nl
GJ 884	&	23 00 16.12	&	-22 31 27.65	&	-904	&	58	&	7.89	&	4.48	&	K5V	&	8.14	&	July 2004	&	450	\nl
GJ 889 A	&	23 07 07.06	&	-23 09 34.01	&	154	&	-254	&	9.61	&	6.42	&	K6V	&	21.11	&	July 2004	&	650	\nl
GJ 892	&	23 13 16.98	&	+57 10 06.08	&	2075	&	295	&	5.56	&	3.26	&	K3V	&	6.53	&	June 2004	&	5.4	\nl
GJ 898	&	23 32 49.40	&	-16 50 44.31	&	344	&	-218	&	8.60	&	5.47	&	K6Vk	&	13.95	&	July 2004	&	900	\nl
\enddata
\end{deluxetable}

\begin{deluxetable}{lc}
\tablecaption{Table of Observations \label{obs}}	
\tablehead{\colhead{Date}	&	\colhead{Seeing Conditions ['']$^a$}}		
\startdata
1-3 June 2004 & 0.5 \nl
1-3 Aug 2004 & 1.0\nl
6-7 Oct 2004 & 0.5 \nl
11 - 14 Nov 2004 &  0.7 \nl
31 Dec 2004 & 2.0\nl
7 Jul 2005 & 1.5\nl
12-14 Nov 2005 & 0.7 \nl
7 Dec 2005 & 3.0 \nl
\enddata
\tablenotetext{a}{Seeing is estimated from the FWHM of one of the target stars observed with the adaptive optics turned off.}
\end{deluxetable}	

\begin{deluxetable}{ll}	
\tablecaption{Stars with No Companion Candidates in the Field of View\label{nondet}}
\tablehead{	\colhead{}								
}	
\startdata									
GJ 10	&	GJ 628	 \nl
GJ 34 A	&	GJ 629.1	 \nl
GJ 34 B	&	GJ 663 A   \nl
GJ 37	&	GJ 663 B	 \nl
GJ 61 A	&	GJ 664	 \nl
GJ 68	&      GJ 670 B	 \nl
GJ 71	&	GJ 678 A	\nl
GJ 79	&	GJ 687	\nl
GJ 107 A	&	GJ 692	\nl
GJ 111	&	GJ 699	\nl
GJ 137	&	GJ 702 B  \nl
GJ 147	&	GJ 725 A  \nl
GJ 167	&	GJ 725 B  \nl
GJ 204 	&	GJ 768	\nl
GJ 303	&	GJ 785     \nl
GJ 338 A	&	GJ 796	\nl
GJ 382	&	GJ 805     \nl
GJ 394	&	GJ 811     \nl
GJ 423.1	&      GJ 820 A  \nl
GJ 447	&      GJ 821	\nl
GJ 448	&	GJ 849	\nl
GJ 506 	&	GJ 872 A	\nl
GJ 526	&	GJ 875	\nl
GJ 527 A	&	GJ 876	\nl 
GJ 555	&	GJ 882	\nl
GJ 557	&	GJ 884	\nl
GJ 570 A	&	GJ 889 A	\nl
GJ 581	&	GJ 898	\nl
GJ 598	&   	GJ 3175	\nl
GJ 602	&	GJ 4324   \nl
GJ 606.2	&	GJ 9207	\nl
GJ 616	&	GJ 9491	\nl
\enddata									
\end{deluxetable}	

\begin{deluxetable}{lcrrccc}
\tablecaption{Companion Candidates \label{comptab}}
\tablehead{ \colhead{Target}  & \colhead{Ks$_{target}$$^a$} & \colhead{Separation}  & \colhead{PA}           &  \colhead{ Ks$_{comp}$} & \colhead{Epoch} & \colhead{Status$^b$} \nl
                                                     &                                              &\colhead{"}                     & \colhead{deg         } &                                                              &  &}
\startdata
GJ 15 A	&	4.02$\pm$0.02	&	6.40$\pm$0.12	&	158.3$\pm$3.2&	15.98$\pm$0.06	&	Oct 2004	&	NCPM	\nl
GJ 15 B	&	5.95$\pm$0.02	&	7.54$\pm$0.12	&	66.2$\pm$0.4	&	16.52$\pm$0.06	&	Oct 2004	&	NCPM	\nl
                   &	                       	&	11.91$\pm$0.14&	70.3$\pm$0.3	&	16.20$\pm$0.06	&	Oct 2004	&	NCPM	\nl
GJ 72	&	4.60$\pm$0.02	&	9.93$\pm$0.12	&	-79.4$\pm$0.4	&	11.85$\pm$0.06	&	Aug 2004	&	NCPM	\nl
GJ 105 A	&	3.48$\pm$0.21	&	2.63$\pm$0.12	&	-54$\pm$0.2	&	8.77$\pm$0.22	&	Aug 2004	&	CPM$^c$	\nl
GJ 107 B	&	5.87$\pm$0.02	&	6.24$pm$0.12	&	161.3$\pm$4.3	&	17.70$\pm$0.06	&	Oct 2004	&	U	\nl
GJ 115 A	&	6.70$\pm$0.02	&	6.50$\pm$0.12	&	156.2$\pm$2.5	&	18.81$\pm$0.06	&	Aug 2004	&	U	\nl
GJ 250 B	&	5.72$\pm$0.04	&	9.62$\pm$0.13	&	-27.2$\pm$1.4	&	14.04$\pm$0.07	&	Nov 2005	&	U	\nl
GJ 302	&	4.17$\pm$0.04	&	9.99$\pm$0.13	&	-152.5$\pm$1.3	&	15.46$\pm$0.07	&	Dec 2004	&	NCPM	\nl
GJ 454	&	4.03$\pm$0.26	&	0.99$\pm$0.12	&	-75.5$\pm$03.2	&	7.75$\pm$0.06	&	July 2005	&	CPM	\nl
GJ 670 A	&	3.07$\pm$0.30	&	10.79$\pm$0.14	&	-98.9$\pm$0.4	&	13.74$\pm$0.38	&	June 2004	&	U	\nl
GJ 701	&	5.31$\pm$0.02	&	12.36$\pm$0.13	&	136.2$\pm$0.3	&	...	&	Aug 2004	&	NCPM	\nl
GJ 702 AB	&	1.79$\pm$0.30	&	9.38$\pm$0.13	&	-128.9$\pm$0.3	&	13.01$\pm$0.30	&	Aug 2004	&	U	\nl
	                  &		&	11.53$\pm$0.14	&	-164.5$\pm$0.3	&	13.82$\pm$0.30	&	Aug 2004	&	U	\nl
GJ 716	&	4.70$\pm$0.02	&	2.84$\pm$0.12	&	170.0$\pm$3.2	&	14.06$\pm$0.06	&	Aug 2004	&	NCPM	\nl
	&		&	3.54$\pm$0.13	&	-92.8$\pm$0.9	&	15.72$\pm$0.06	&	Aug 2004	&	NCPM	\nl
	&		&	5.11$\pm$0.13	&	-147.4$\pm$1.4	&	15.75$\pm$0.06	&	Aug 2004	&	NCPM	\nl
	&		&	5.30$\pm$0.14	&	71.2$\pm$1.2	&	15.91$\pm$0.06	&	Aug 2004	&	NCPM	\nl
	&		&	5.38$\pm$0.13	&	119.0$\pm$1.6	&	13.62$\pm$0.06	&	Aug 2004	&	NCPM	\nl
GJ 722$^d$	&	4.23$\pm$0.02	&	5.02$\pm$0.12	&	-20.2$\pm$5.4	&	15.90$\pm$0.06	&	Aug 2004	&	NCPM	\nl
	&		&	6.40$\pm$0.12	&	30.1$\pm$1.5	&	17.50$\pm$0.06	&	Aug 2004	&	NCPM	\nl
GJ 726$^d$	&	5.58$\pm$0.03	&	2.11$\pm$0.12	&	-93.6$\pm$1.5	&	13.63$\pm$0.06	&	June 2004	&	U	\nl
	&		&	2.46$\pm$0.12	&	-136.0$\pm$1.4	&	14.31$\pm$0.06	&	June 2004	&	U	\nl
		&		&	3.09$\pm$0.12	&	-122.7$\pm$1.0	&	17.62$\pm$0.06	&	June 2004	&	U	\nl
	&		&	3.85$\pm$0.12	&	-4.8$\pm$1.2	&	16.42$\pm$0.06	&	June 2004	&	U	\nl
GJ 729	&	5.37$\pm$0.02	&	6.24$\pm$0.12	&	-49$\pm$0.5	&	17.04$\pm$0.06	&	Aug 2004	&	U	\nl
	&		&	8.34$\pm$0.13	&	-126.0$\pm$0.4	&	17.70$\pm$0.06	&	Aug 2004	&	NCPM	\nl
GJ 779	&	4.39$\pm$0.03	&	3.30$\pm$0.12	&	-77.5$\pm$0.9	&	13.53$\pm$0.06	&	Aug 2004	&	NCPM	\nl
	&		&	6.21$\pm$0.12	&	162.0$\pm$5.2	&	12.38$\pm$0.06	&	Aug 2004	&	NCPM	\nl
	&		&	9.34$\pm$0.13	&	140.1$\pm$0.9	&	12.97$\pm$0.06	&	Aug 2004	&	NCPM	\nl
GJ 789	&	4.90$\pm$0.02	&	12.78$\pm$0.14	&	67.9$\pm$0.3	&	15.41$\pm$0.06	&	Aug 2004	&	U	\nl
GJ 820 B	&	2.54$\pm$0.33	&	11.51$\pm$0.14	&	77.4$\pm$0.4	&	14.8$\pm$0.06	&	Nov 2005	&	U	\nl
GJ 848.4	&	4.60$\pm$0.02	&	6.69$\pm$0.12	&	-158.2$\pm$3.1	&	16.5$\pm$0.06	&	June 2004	&	U	\nl
GJ 873	&	5.30$\pm$0.02	&	6.47$\pm$0.12	&	175.5$\pm$3.0	&	...	&	Nov 2005	&	U	\nl
	&		&	5.78$\pm$0.12	&	115.6$\pm$0.2	&	...	&	Nov 2005	&	U	\nl
GJ 892	&	3.26$\pm$0.30	&	6.99$\pm$0.12	&	44.5$\pm$0.5	&	...	&	June 2004	&	NCPM	\nl
	&		&	9.93$\pm$0.13	&	-17.9$\pm$0.9	&	...	&	June 2004	&	U	\nl
	&		&	11.14$\pm$0.12	&	-74.4$\pm$0.8	&	...	&	June 2004	&	U	\nl
GJ 1020	&	4.62$\pm$0.02	&	3.98$\pm$0.12	&	-119.4$\pm$0.7	&	10.34$\pm$0.06	&	Dec 2004	&	CPM	\nl
\enddata
\tablenotetext{a}{K$_s$ magnitudes and errors taken from the 2MASS catalog (Cutri et al. 2006).}
\tablenotetext{b}{NCPM= non-common proper motion, CPM=common proper motion, U=unconfirmed}
\tablenotetext{c}{Companion around GJ 105a was originally identified by Golimowski et al. 1995}
\tablenotetext{d}{Star in crowded field. Not all companions in the field of view are listed.}
\end{deluxetable}

\begin{deluxetable}{lcrrccc}
\tablecaption{Astrometry for Confirmed Companions \label{compastron}}
\tablehead{ \colhead{Target}  & \colhead{$\rho$ [$\arcsec$]} & \colhead{PA [$\deg$]}  & \colhead{Epoch} }
\startdata
GL 454	&	0.987$\pm$0.115	         &	-75.06$\pm$3.23	& June 4 2004	\nl
                   &       0.992$\pm$0.115            &      -75.51$\pm$3.12  & July 7 2005      \nl
GL 1020	&	3.98$\pm$0.12	         &	-119.35$\pm$0.74	& Dec 31 2004	\nl
                   &       4.01$\pm$0.12         &        -119.97$\pm$0.74   &  Nov 14 2005      \nl
\enddata
\end{deluxetable}

\begin{deluxetable}{lcccc}								
\tablecaption{Palomar Imaging Sensitivities, $\Delta$K$_S$ at 90\% completeness \label{sentab}}	
\tablehead{ \colhead{Target} & \colhead{0.5''} & \colhead{1''} & \colhead{2''} & \colhead{5''} 
}											
\startdata	
GJ 10	&	6.32	&	6.54	&	7.91	&	10.20	\nl
GJ 1020	&	6.08	&	6.01	&	7.16	&	9.30	\nl
GJ 105a	&	5.82	&	5.93	&	7.25	&	9.29	\nl
GJ 107a	&	6.04	&	6.18	&	7.61	&	10.43	\nl
GJ 107b	&	7.52	&	8.33	&	10.21	&	12.77	\nl
GJ 111	&	6.31	&	6.44	&	8.02	&	9.74	\nl
GJ 115a	&	9.67	&	9.92	&	11.66	&	13.46	\nl
GJ 137	&	5.45	&	5.70	&	7.04	&	8.96	\nl
GJ 147	&	5.73	&	6.01	&	7.56	&	10.24	\nl
GJ 15a	&	6.61	&	6.95	&	7.99	&	10.24	\nl
GJ 15b	&	8.00	&	8.26	&	9.55	&	12.22	\nl
GJ 204	&	6.95	&	7.17	&	8.44	&	10.30	\nl
GJ 303	&	6.54	&	6.97	&	8.36	&	10.54	\nl
GJ 3175	&	6.86	&	6.98	&	8.48	&	10.20	\nl
GJ 37	&	6.16	&	6.25	&	8.06	&	9.95	\nl
GJ 454	&	5.56	&	5.79	&	7.35	&	8.45	\nl
GJ 506	&	5.24	&	5.36	&	6.23	&	8.54	\nl
GJ 527a	&	6.37	&	6.10	&	7.51	&	9.37	\nl
GJ 602	&	4.55	&	4.82	&	5.78	&	7.26	\nl
GJ 606.2	&	5.62	&	5.40	&	7.15	&	8.76	\nl
GJ 616	&	5.91	&	5.88	&	7.55	&	8.25	\nl
GJ 61a	&	7.60	&	8.25	&	9.75	&	12.21	\nl
GJ 629.1	&	9.56	&	8.64	&	10.02	&	10.93	\nl
GJ 68	&	6.04	&	6.33	&	7.83	&	9.79	\nl
GJ 699	&	6.19	&	6.54	&	8.60	&	11.08	\nl
GJ 702a	&	4.34	&	4.87	&	6.25	&	8.97	\nl
GJ 71	&	7.81	&	4.73	&	5.43	&	8.50	\nl
GJ 725b	&	6.57	&	6.94	&	8.08	&	9.72	\nl
GJ 729	&	7.51	&	7.72	&	8.79	&	9.97	\nl
GJ 779	&	6.98	&	6.97	&	8.81	&	10.38	\nl
GJ 785	&	5.98	&	6.19	&	7.64	&	10.29	\nl
GJ 789	&	7.31	&	7.57	&	6.84	&	9.18	\nl
GJ 805	&	5.70	&	5.72	&	6.84	&	9.18	\nl
GJ 811	&	6.62	&	6.61	&	7.51	&	9.25	\nl
GJ 820a	&	3.67	&	3.95	&	4.94	&	7.36	\nl
GJ 820b	&	4.44	&	4.54	&	5.75	&	7.78	\nl
GJ 849	&	7.84	&	8.11	&	9.70	&	12.00	\nl
GJ 872a	&	5.67	&	5.71	&	7.43	&	9.15	\nl
GJ 873	&	7.68	&	7.99	&	8.96	&	11.07	\nl
GJ 882	&	6.32	&	6.21	&	8.12	&	9.90	\nl
GJ 884	&	7.45	&	7.66	&	8.95	&	11.14	\nl
GJ 889a	&	8.18	&	7.92	&	9.05	&	9.45	\nl
GJ 898	&	9.54	&	9.91	&	10.00	&	12.35	\nl
GJ 9491	&	6.20	&	6.28	&	7.73	&	10.82	\nl										
\enddata											
\end{deluxetable}

\begin{figure}[ht]
\figcaption{Images of the SIM Lite target stars with unidentified or confirmed background stars in their 25$\arcsec$ field-of-view. We have circled some of the stars to aid the reader. (These figures are available upon request.) 
\label{plotstars}}
\end{figure}
\begin{figure}[ht]
\end{figure}
\begin{figure}[ht]
\end{figure}
\begin{figure}[ht]
\end{figure}
\begin{figure}[ht]
\end{figure}

\end{document}